\documentclass{elsart}

 \usepackage{graphicx}
\usepackage{amssymb}

\begin{document}

\begin{frontmatter}

% Title, authors and addresses

% use the thanksref command within \title, \author or \address for footnotes;
% use the corauthref command within \author for corresponding author footnotes;
% use the ead command for the email address,
% and the form \ead[url] for the home page:
% \title{Title\thanksref{label1}}
% \thanks[label1]{}
% \author{Name\corauthref{cor1}\thanksref{label2}}
% \ead{email address}
% \ead[url]{home page}
% \thanks[label2]{}
% \corauth[cor1]{}
% \address{Address\thanksref{label3}}
% \thanks[label3]{}

\title{The Origin and Control of the Sources of AMR in (Ga,Mn)As Devices}

% use optional labels to link authors explicitly to addresses:
% \author[label1,label2]{}
% \address[label1]{}
% \address[label2]{}

\author[nott]{A.~W.~Rushforth\corauthref{cor}},
\corauth[cor]{Corresponding author.}
\ead{Andrew.rushforth@nottingham.ac.uk}
\author[prag]{K.~V\'yborn\'y},
\author[nott]{C.~S.~King},
\author[nott]{K.~W.~Edmonds},
\author[nott]{R.~P.~Campion},
\author[nott]{C.~T.~Foxon},
\author[hit]{J.~Wunderlich},
\author[mrc]{A.~C.~Irvine},
\author[prag]{V.~Nov\'ak},
\author[prag]{K.~Olejn\'{\i}k},
\author[texas]{A.~A.~Kovalev},
\author[texas]{Jairo~Sinova},
\author[prag,nott]{T.~Jungwirth}, and
\author[nott]{B.~L.~Gallagher}

%\author{}
\address[nott]{School of Physics and Astronomy, University of Nottingham, 
Nottingham~NG7~2RD, UK}
\address[prag]{Institute of Physics, 
Academy of Sciences of the Czech Republic, 
Cukrovarnick\'a~10, 162 53 Praha 6, Czech Republic}
\address[hit]{Hitachi Cambridge Laboratory, Cambridge CB3 0HE, UK}
\address[mrc]{Microelectronics Research Centre, Cavendish Laboratory, 
University of Cambridge, CB3 0HE, UK}
\address[texas]{Department of Physics, Texas A\&M University, College
Station, TX 77843-4242, USA}
\address{}

\begin{abstract}
% Text of abstract
  We present details of our experimental and theoretical study of the
  components of the anisotropic magnetoresistance (AMR) in (Ga,Mn)As. We
  develop experimental methods to yield directly the non-crystalline and
  crystalline AMR components which are then independently analyzed. These
  methods are used to explore the unusual phenomenology of the AMR in ultra
  thin (5nm) (Ga,Mn)As layers and to demonstrate how the components of the AMR
  can be engineered through lithography induced local lattice relaxations. We
  expand on our previous [Phys. Rev. Lett. \textbf{99}, 147207 (2007)] theoretical analysis and
  numerical calculations to present a simplified analytical model for the
  origin of the non-crystalline AMR. We find that the sign of the
  non-crystalline AMR is determined by the form of spin-orbit coupling in the
  host band and by the relative strengths of the non-magnetic and magnetic
  contributions to the impurity potential.
%We discuss generic implications of our experimental and theoretical findings 
%including predictions for
%non-crystalline AMR sign reversals in dilute moment systems.

\end{abstract}

\end{frontmatter}

% main text
\section{Introduction}
\label{sec1}

The Anisotropic Magnetoresistance (AMR) in ferromagnetic metals has been the
subject of many studies since it was discovered more than a century
ago\cite{Thomson}. The effect describes the change of the electrical
resistance in response to a change in the orientation of the magnetization.
Phenomenologically, AMR has a non-crystalline component, arising from the
lower symmetry for a specific current direction, and crystalline components
arising from the crystal symmetries \cite{Doring:1938_a,vanGorkom:2001_a}. In
ferromagnetic metals the AMR components have been extracted indirectly from
experimental data by fitting the total angular dependences
\cite{vanGorkom:2001_a}. The coefficients can be obtained by numerical {\em ab
  initio} transport calculations \cite{Banhart:1995_a}, but these have no
clear connection to the standard physical model of transport arising from spin
dependent scattering of current carrying low mass $s$-states into heavy-mass
$d$-states \cite{McGuire:1975_a}.

In the dilute magnetic semiconductor (Ga,Mn)As, it has been observed that the
AMR consists of a non-crystalline component of the opposite sign (compared to
most ferromagnetic metals) and typically much weaker crystalline terms
reflecting the underlying magnetocrystalline
anisotropies \cite{Baxter:2002_a,Jungwirth:2003_b,Tang:2003_a,Matsukura:2004_a,Goennenwein:2004_a,Wang:2005_c,Limmer:2006_a}.
In our recent work \cite{Rushforth:2007} we developed experimental methods to
yield directly the non-crystalline and crystalline components of the AMR in
(Ga,Mn)As. Numerical microscopic calculations were employed to calculate the
AMR components, achieving good agreement with experimental results. An
analytical model was introduced to explain the physical origin of the AMR
terms, in particular to show that the sign of the non-crystalline term is
determined by the ratio of the magnetic and non-magnetic components of the
scattering potential. Here we expand our previous work to include details of
that analytical model and a detailed explanation of how the components of the
AMR can be separately extracted from the experimental data.

\section{Experiment}
\label{sec2}

25nm and 5nm Ga$_{0.95}$Mn$_{0.05}$As films were grown by low temperature
molecular beam epitaxy onto a 330nm HT-GaAs buffer layer grown on a
semi-insulating GaAs(001) substrate. Standard photolithography techniques were
used to fabricate two sets of Hall bars of width 45$\mu$m with voltage probes
separated by 285$\mu$m, with the current along the [100], [010], [110] and
$[1\bar{1}0]$ directions. The longitudinal resistance R$_{xx}$ and the Hall
resistance R$_{xy}$ were measured using four probe techniques with a DC
current of 1$\mu$A. We also measured devices with a Corbino geometry (Fig.
1(d)) in which the current flows radially in the plane of the material. All measurements were made on as-grown material (i.e. without annealing). 
Magnetoresistances were measured with a saturating magnetic field of 1T
applied in the plane of the device, {\em i.e.}, in the pure AMR geometry with
zero (antisymmetric) Hall signal. Magnetometry measurements were carried out
using a commercial Quantum Design SQUID magnetometer.

\begin{figure}
\hspace*{0cm}\includegraphics[width=140mm,angle=-0]{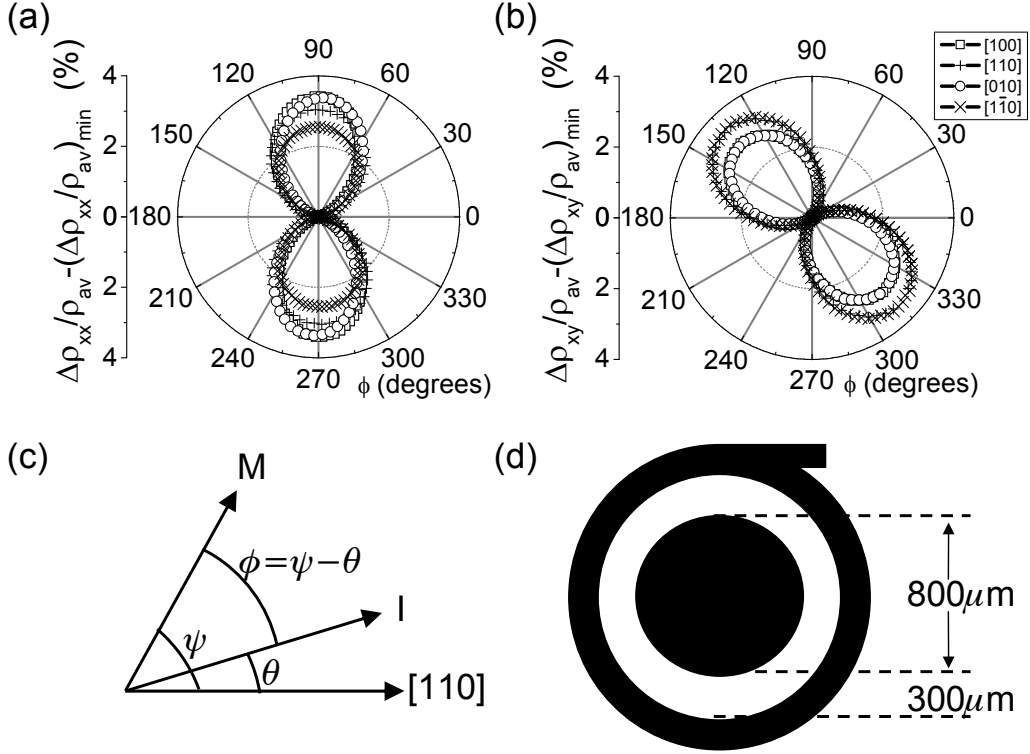}
\caption{(a) 
  Measured (at 4.2~K) longitudinal and (b) transverse AMR for as-grown 25nm
  thick Ga$_{0.95}$Mn$_{0.05}$As film as a function of the angle between
  magnetization in the plane of the film and the current. The legend shows the
  direction of the current. The y-axes show $\Delta\rho/\rho_{av}$ shifted so
  that the minimum is at zero, to show the symmetries present in the data more
  clearly. (c) Definition of the angles referred to in the text. (d) Corbino
  geometry.}
\label{1}
\end{figure}

Figures 1(a) and (b) show the fractional change in the resistivity,
$\Delta\rho_{xx}/\rho_{av}$ and $\Delta\rho_{xy}/\rho_{av}$
($\Delta\rho_{xx}=\rho_{xx}-\rho_{av}$) for the Hall bars fabricated from the
25nm film for each current direction. Here $\rho_{av}$ is the average value of
the longitudinal resistance as the magnetic field is rotated through
$360^\circ$. In the following analysis we will show that this data can be
decomposed into the form given by equations (\ref{rho_xx}) and
(\ref{rho_xy}). In the subsequent sections we aim to identify the origin of the
particular contributions.

The terms allowed by
symmetry are obtained by extending the standard phenomenology
\cite{Doring:1938_a}, to systems with cubic [100] plus uniaxial [110]
anisotropy:
\begin{eqnarray}
\frac{\Delta\rho_{xx}}{\rho_{av}} =C_I\cos2\phi + C_U\cos2\psi + C_C\cos4\psi
 + C_{I,C}\cos(4\psi-2\phi)\;.
\label{rho_xx}
\end{eqnarray}
\begin{equation}
\frac{\Delta\rho_{xy}}{\rho_{av}}
=C_I\sin2\phi - C_{I,C}\sin(4\psi-2\phi)\; .
\label{rho_xy}
\end{equation}
$\phi$ is the angle between the magnetization unit vector ${\hat{M}}$ and
the current $I$, and $\psi$ the angle between ${\hat{M}}$ and the
[110] crystal direction.  The four contributions are the non-crystalline term,
the lowest order uniaxial and cubic crystalline terms, and a crossed
non-crystalline/crystalline term. We have omitted the higher order terms in
these expressions as these are found to be negligibly small in our devices.
The purely crystalline terms are excluded by symmetry for the transverse AMR
and it is clear that $\Delta\rho_{xy}/\rho_{av}$ and
$\Delta\rho_{xx}/\rho_{av}$ are not independent. The well known formulae for
isotropic materials (see {\em e.g.}  \cite{Tang:2003_a}) correspond to
$C_I\not=0$ and all other coefficients set to zero in equations
(\ref{rho_xx},\ref{rho_xy}). Such formulae result also from equations
(\ref{rho_xx},\ref{rho_xy}) averaged over $\psi$ with $\phi$ held constant,
which corresponds to a polycrystal with randomly oriented grains.

It is not possible to obtain each term separately by fitting equations
(\ref{rho_xx}) and (\ref{rho_xy}) to the full ($\psi$) angular data from
one Hall bar. In a $[1\bar{1}0]$ oriented Hall bar, for instance,
$\phi=\psi+\frac{\pi}{2}$, the equations reduce to $\Delta\rho_{xx}/\rho_{av}=
(-C_I+C_U-C_{I,C})\cos 2\psi + C_C\cos 4\psi$ and
$\rho_{xy}/\rho_{av}=-C_I+C_{I,C}\sin 2\psi$. $C_I$ and $C_{I,C}$ could be
determined only if $C_U=0$. However, using different Hall bar orientations,
the anisotropy constants appear in other combinations so that they can all be
extracted individually. In the present case, two Hall bar orientations (e.g.
$[110]$ and $[1\bar{1}0]$) suffice. Measurements with other orientations are
used as consistency checks.

Figure 2(a) shows some examples of how the individual AMR components can be
extracted: the longitudinal resistivities in the $[010]$ and $[100]$
directions are subtracted to give a simple $\sin 2\psi$ signal
(cf.~Eq.~\ref{rho_xx}) with amplitude $2(C_I-C_{I,C})$ and this is the same as
the difference of transverse resistivities in the $[110]$ and $[1\bar{1}0]$
samples. A $\cos 2\psi$ signal with amplitude $2(C_I+C_{I,C})$ can be obtained
through other combinations of the data. The close agreement obtained by using
different combinations to extract the same coefficients highlights the
consistency between the data and our phenomenological analysis, and also
attests to the homogeneity and quality of the MBE grown GaMnAs wafer.

The purely crystalline terms can be extracted from the combination of the
measurements from two orthogonal bars. This is shown in Fig. 2(b) for our
data. In principle, the same signals could be extracted directly from a single
experimental measurement by appropriate wiring of an $L$-shaped Hall bar.
Alternatively, the same coefficients can be extracted directly from the data
obtained from the Corbino geometry, Fig. 2(b). As the current flows in all
radial directions, the non-crystalline terms are averaged to
zero\cite{Corbino}. It would be problematic to perform such measurements with
ferromagnetic metals because, in order to define the Corbino geometry it is
required that the source-drain resistance be large compared to the contact
resistances. This condition is met with (Ga,Mn)As. Again, we obtain excellent
agreement for the same coefficients extracted from different combinations of
devices.

% With appropriate wiring on
%two $L$-shaped Hall bars such two signals can be directly the experimental
%output giving a stringent consistency check and requiring only very
%transparent fitting if values of anisotropy constants are needed. The two
%overlapping data traces of filled symbols in Fig.~\ref{2}a were obtained
%by directly subtracting the raw experimental data from different samples and
%thus attest the constant sample properties across the batch. Another such
%consistency check is explained in Fig. 2(a) (empty symbols).

\begin{figure}
\hspace*{0cm}\includegraphics[width=140mm,angle=-0]{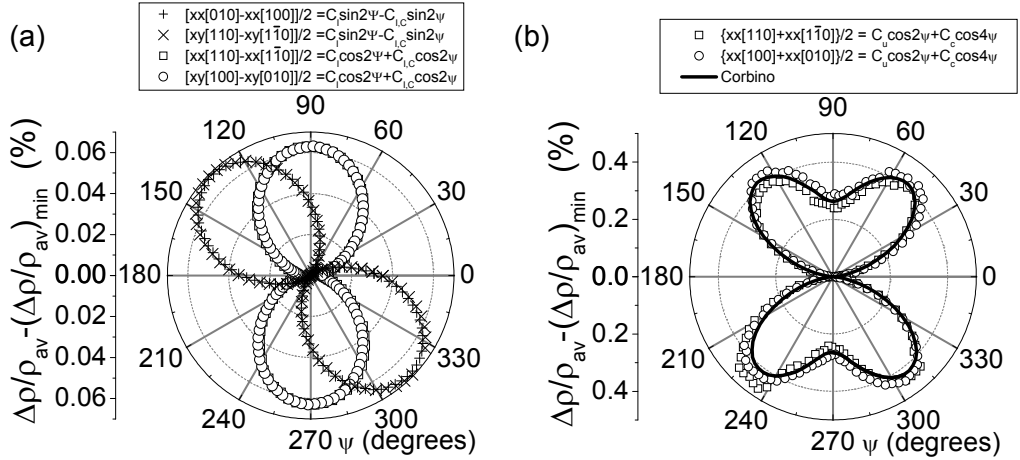}
\caption{Combinations of the data from Fig.1 
  used to extract the AMR components. The notation xx[100] is used to denote
  $\Delta\rho_{xx}/\rho_{av}$ for current along the [100] direction etc. }
\label{fig2}
\end{figure}

Figure 3(a) shows the coefficients C$_{I,C}$, C$_{U}$ and C$_{C}$ extracted
from the Hall bar and Corbino disk data over the whole range up to the Curie
temperature (80K). Note that the uniaxial crystalline term, C$_{U}$, becomes
the dominant term for T$\geqslant$30K. This correlates with the uniaxial
component of the magnetic anisotropy which dominates for T$\geqslant$30K as
observed by SQUID magnetometry measurements (Figure 3(b)). Our work shows that
in (Ga,Mn)As ferromagnets, the symmetry breaking mechanism behind this
previously reported \cite{Sawicki:2004_a} uniaxial magneto-crystalline
anisotropy in the magnetization also contributes to the AMR.

\begin{figure}
\hspace*{0cm}\includegraphics[width=140mm,angle=-0]{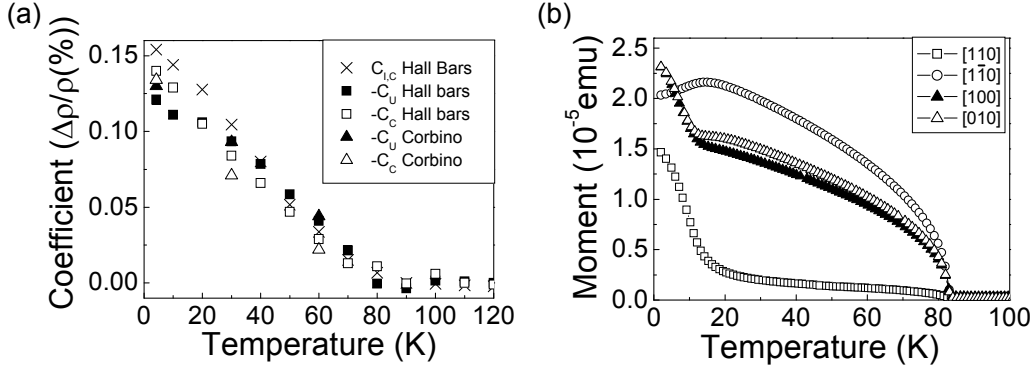}
\caption{(a) Temperature 
  dependence of the crystalline terms extracted from the Hall bars and Corbino
  devices. (b) Remnant moment of 25nm Ga$_{0.95}$Mn$_{0.05}$As film measured
  by SQUID magnetometer along different crystalline directions after cooling
  in a field of 0.1T.}
\label{fig3}
\end{figure}

In our previous work \cite{Rushforth:2007} we used the techniques explained
above to extract the individual AMR components for ultra thin (5nm)
Ga$_{0.95}$Mn$_{0.05}$As films for which the AMR (Fig. 4(a)) is very different
from that observed in the 25nm film. For the 5nm film, the uniaxial
crystalline term dominates the AMR as shown in the Fig. 4(b) for the device in
the Corbino geometry. These techniques were also used to demonstrate how the
crystalline terms can be tuned by the use of lithographic patterning to induce
an additional uniaxial anisotropy in very narrow Hall bars. It has been
shown \cite{Humpfner,Wunderlich} that the fabrication of narrow bars allows the
in-plane compressive strain in the (Ga,Mn)As film to relax in the direction
along the width of the Hall bar and this can lead to an additional uniaxial
component in the magnetocrystalline anisotropy for bars with widths on the
order of 1$\mu$m or smaller. Figs.~5(a) and (b) show the AMR of 45$\mu$m wide
bars and 1$\mu$m wide bars fabricated from nominally identical 25nm
Ga$_{0.95}$Mn$_{0.05}$As wafers. For the 45$\mu$m bars, the cubic crystalline
symmetry leads to the AMR along [100] and [010] being larger than along [110]
and [1$\bar{1}$0]. For the narrow bars we observe the opposite relationship,
consistent with the addition of an extra uniaxial component which adds 0.8\%
to the AMR when current is along [110] and [1$\bar{1}$0] and subtracts 0.4\%
when the current is along [100] and [010].

It is anticipated that the experimental techniques developed here will be useful in studying the magnetotransport coefficients in other magnetic materials and
nanostructures.

\begin{figure}
\hspace*{0cm}\includegraphics[width=140mm,angle=-0]{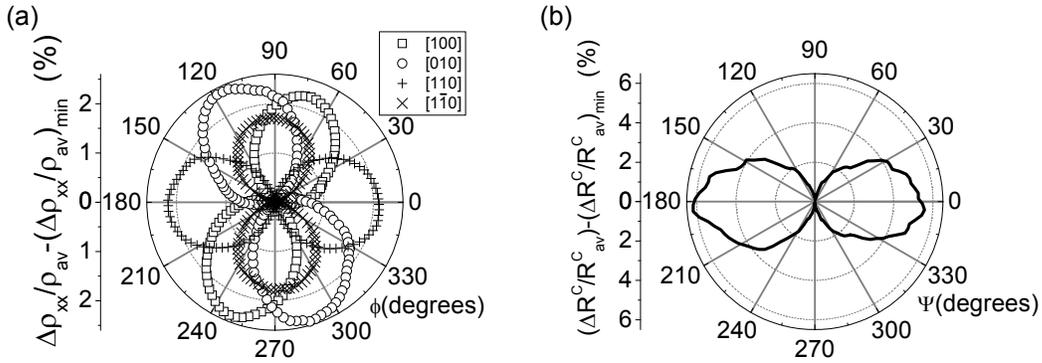}
\caption {Longitudinal AMR 
  of the 5nm Ga$_{0.95}$Mn$_{0.05}$As Hall bars. T=20K. (b) AMR of a 5nm
  Ga$_{0.95}$Mn$_{0.05}$As film in the Corbino geometry. T=11K.}
\label{fig4}
\end{figure}

%The 5nm films have lower Curie temperatures ($T_{C}\approx$30K) than the 25nm
%films and become highly resistive at low temperature indicating that they are
%close to the metal-insulator transition. (The 25nm films show metallic
%behaviour down to the lowest measured temperatures.) The unusual behaviour of
%the AMR in the 5~nm films is not reproduced by theory simulations assuming
%weakly disordered, fully delocalized (Ga,Mn)As valence bands (see later). It
%might be related to the expectation that magnetic interactions become more
%anisotropic with increasing localization of the holes near their parent Mn
%ions as the metal-insulator transition is approached~\cite{Jungwirth:2006_a}.

\begin{figure}[strain]
\hspace*{0cm}\includegraphics[width=140mm,angle=-0]{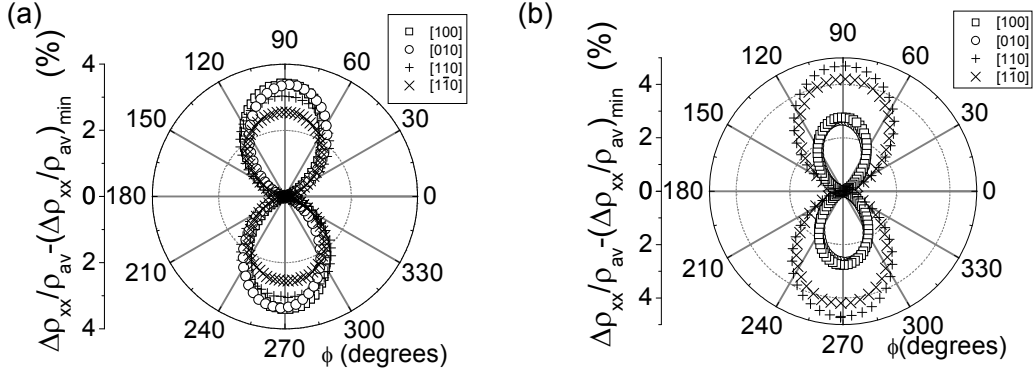}
\caption {(a) AMR for macroscopic Hall bars and (b) 
  narrow (1$\mu$m wide) Hall bars fabricated from 25nm
  Ga$_{0.95}$Mn$_{0.05}$As films. T=4.2K.}
\label{fig5}
\end{figure}

%Finally, we demonstrate how the crystalline terms can be tuned by the use of
%lithographic patterning to induce an additional uniaxial anisotropy in very
%narrow Hall bars. In recent studies \cite{Humpfner,Wunderlich} it has been
%found that the patterning allows the in-plane compressive strain in the
%(Ga,Mn)As film to relax in the direction along the width of the Hall bar and
%this can lead to an additional uniaxial component in the magnetocrystalline
%anisotropy for bars with widths on the order of 1$\mu$m or smaller. Figs.~5(a)
%and (b) show the AMR of 45$\mu$m wide bars and 1$\mu$m wide bars fabricated
%from nominally identical 25nm Ga$_{0.95}$Mn$_{0.05}$As wafers. For the
%45$\mu$m bars, the cubic crystalline symmetry leads to the AMR along [100] and
%[010] being larger than along [110] and [1$\bar{1}$0]. For the narrow bars we
%observe the opposite relationship. This is consistent with the addition of an
%extra uniaxial component, whose presence in the magnetocrystalline anisotropy
%is confirmed by SQUID magnetometry measurements (not shown), which adds 0.8\%
%to the AMR when current is along [110] and [1$\bar{1}$0] and subtracts 0.4\%
%when the current is along [100] and [010]. These post-growth lithography
%induced modifications are significant fractions of the total AMR of the parent
%(Ga,Mn)As material.

\section{From the full Boltzmann theory simulations to a simple analytical
  model of the AMR}
\label{sec3}

\begin{figure}[]
\hspace*{0cm}\includegraphics[width=140mm,angle=-0]{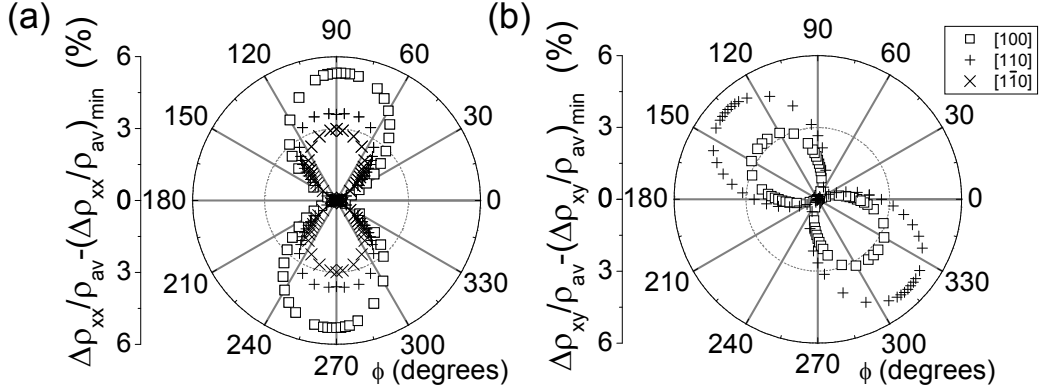}
\caption{Calculated angular dependence of the AMR (see the text for system
  parameters).}
\label{fig6}
\end{figure}

A six-band $\vec{k}\cdot\vec{p}$ description of the GaAs host valence band
combined with the kinetic-exchange model of the coupling to the local Mn$_{\rm
  Ga}$ $d^5$-moments \cite{Dietl:2001_b} can provide input for the Boltzmann
equation producing the conductivity tensor $\sigma$. In our previous work
\cite{Rushforth:2007}, we expanded the studies of Jungwirth~{\em et~al.}
\cite{Jungwirth:2002_c,Jungwirth:2003_b} by performing numerical calculations
of the $\sigma_{xx}$ for arbitrary in-plane directions of $\hat{M}$ (not just
$\hat{M}||I$ and $\hat{M}\perp I$). The results of the calculations (see
Fig.~\ref{fig6}) are in semi-quantitative agreement with the experimental AMRs
for the 25nm Ga$_{0.95}$Mn$_{0.05}$As films shown in Fig.~\ref{1}~(a,b). The
zero-temperature model of a $15\%$ compensated material \cite{compensation}
shows that both the $\Delta\rho_{xx}/\rho_{\mathrm{av}}$ and
$\Delta\rho_{xy}/\rho_{\mathrm{av}}$ are dominated by the non-crystalline
$\cos 2\phi$ and $\sin 2\phi$ part, being of the order of several percent and
negative ($C_I=-2.2\%$). The crystalline terms are an order of magnitude
smaller ($C_U=-0.2\%$, $|C_C|<0.1\%$, $C_{I,C}=0.4\%$). These calculations
assumed a growth strain $e_0=-0.3\%$ and a uniaxial strain \cite{uniaxialnote}
$e_{xy}=-0.01\%$.

%Expanding the previous studies
%of Jungwirth {\em et al.} \cite{Jungwirth:2002_c,Jungwirth:2003_b} we
%calculated the $\sigma_{xx}$ for arbitrary in-plane directions of $M$ (not
%just $M||I$ and $M\perp I$). Both the $\Delta\rho_{xx}/\rho_{\mathrm{av}}$
%(AMR) and $\Delta\rho_{xy}/\rho_{\mathrm{av}}$ are dominated by the
%non-crystalline $\cos 2\phi$ and $\sin 2\phi$ part being of the order of
%several percent and negative ($C_I<0$, Eq.~\ref{rho_xx}). On this background,
%small crystalline contributions appear, typically an order of magnitude
%smaller than $C_I$. The largest crystalline term is $C_{I,C}$, in line with
%the statement that the amplitude of $\Delta\rho_{xx}(\psi)/\rho_{\mathrm{av}}$
%varies appreciably between different Hall bar orientations, Fig.~\ref{fig6}.

%We rate the agreement to the experimental AMRs for the 25nm
%Ga$_{0.95}$Mn$_{0.05}$As films, Fig.~\ref{1} (a,b), as semi-quantitative. The
%zero-temperature model of a $15\%$ compensated material shows the
%non-crystalline component as the strongest one ($C_I=-2.2\%$), cubic and
%uniaxial crystalline terms\cite{uniaxialnote} are weak but present
%($C_U=0.2\%$, $|C_C|<0.1\%$) and the cross term $C_{I,C}=0.4\%$. These results
%assumed growth strain $e_0=-0.3\%$ and uniaxial strain\cite{uniaxialnote}
%$e_{xy}=-0.01\%$ and are near to the experimental ones.

Previously, we noted that the crystalline terms arise from the warping of the
valence band \cite{Rushforth:2007}. In the rest of this paper we concentrate
only on the non-crystalline component of the AMR ($C_I$) which in our model
originates from anisotropic scattering of spin-orbit coupled holes on Mn$_{\mathrm{Ga}}$
impurities containing polarized local moments.

%Its strength is a
%remarkable feature present across a wide range of bulk systems
%\cite{Jungwirth:2002_c}.  Given its persistence, it is likely that the
%non-crystalline AMR always occurs via a single mechanism and here we propose
%such a mechanism. It is based on anisotropic relaxation rates due to
%anisotropic scattering on Mn$_{\mathrm{Ga}}$ impurities and the physical
%origin of the anisotropy is the polarization of the Mn magnetic moment. Here,
%we will derive in detail the simplified model introduced
%earlier\cite{Rushforth:2007} and its relation to the full model of
%\cite{Jungwirth:2002_c} will be a subject of the next communication.

The six-band Kohn-Luttinger Hamiltonian $H_{KL}$ is parameterized by the
Luttinger parameters $\gamma_1,\gamma_2,\gamma_3$ and SO splitting
$\Delta_{SO}$ while the exchange splitting strength is characterized by the
constant $J_{pd}$ \cite{Dietl:2001_b}. The total Hamiltonian then reads
\begin{equation}\label{eq-01}
  H=H_{KL}+J_{pd}\sum_{i,I} \vec{S}_I\cdot \vec{s}_i 
   \delta(\vec{r}_i-\vec{R}_I)\,,  
\end{equation}
where $\vec{S}_I$ and $\vec{s}_i$ denote the Mn and hole spins placed at
$\vec{r}_i$ and $\vec{R}_I$.  The mean-field treatment of the exchange
splitting (second term in Eq.~(\ref{eq-01})) 
leads to a term formally equivalent
to an effective Zeeman splitting of strength
$h=J_{pd}N_{\mathrm{Mn}}S_{\mathrm{Mn}}$, where $N_{\mathrm{Mn}}$ is the Mn
concentration and $S_{\mathrm{Mn}}=\frac{5}{2}$ for the five Mn $d$-electrons.
In the following we consider the spherical approximation to $H_{KL}$
($\gamma_2=\gamma_3$) and the four-band model
($\Delta_{SO}\rightarrow\infty$). The total Hamiltonian can be then rewritten
as
\begin{equation}\label{eq-02}
  H = \textstyle
    (\gamma_1+\frac{5}{2}\gamma_2) \hbar^2 k^2/2m -
     2\gamma_2 (\vec{k}\cdot \vec{J})^2 \hbar^2/2 m +
    h \hat{M}\cdot \vec{s}\,.
\end{equation}
Explicit expressions of the matrices for the total angular momentum $\vec{J}$
and hole spin $\vec{s}$ can be found in Refs. \cite{Dargys:2004_a} and
\cite{Abolfath:2001_a}.  With this simplified band structure we will now
estimate the AMR assuming that the scattering takes place only on Mn ions
substituting Ga which is the dominant mechanism.

% in high quality annealed
% materials.

\begin{figure}
\begin{center}
\begin{tabular}{cc}
  \includegraphics[scale=0.5]{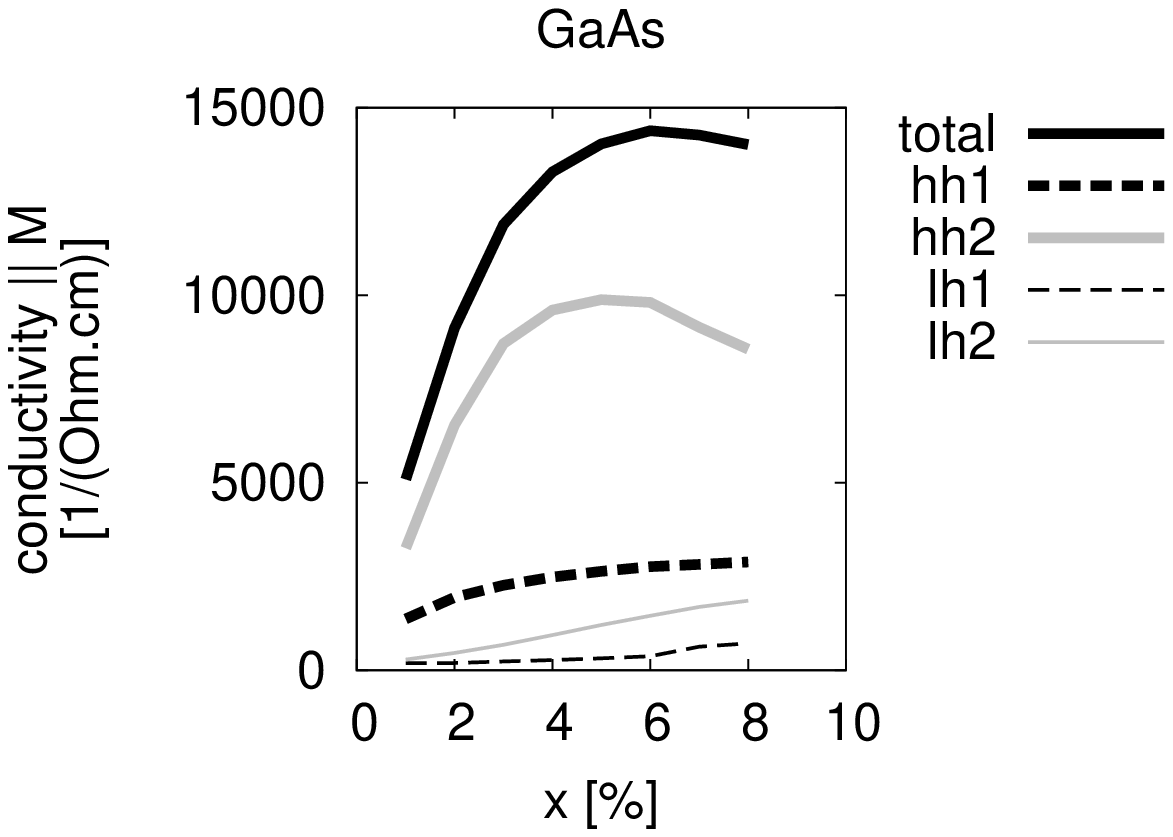} &
  \includegraphics[scale=0.5]{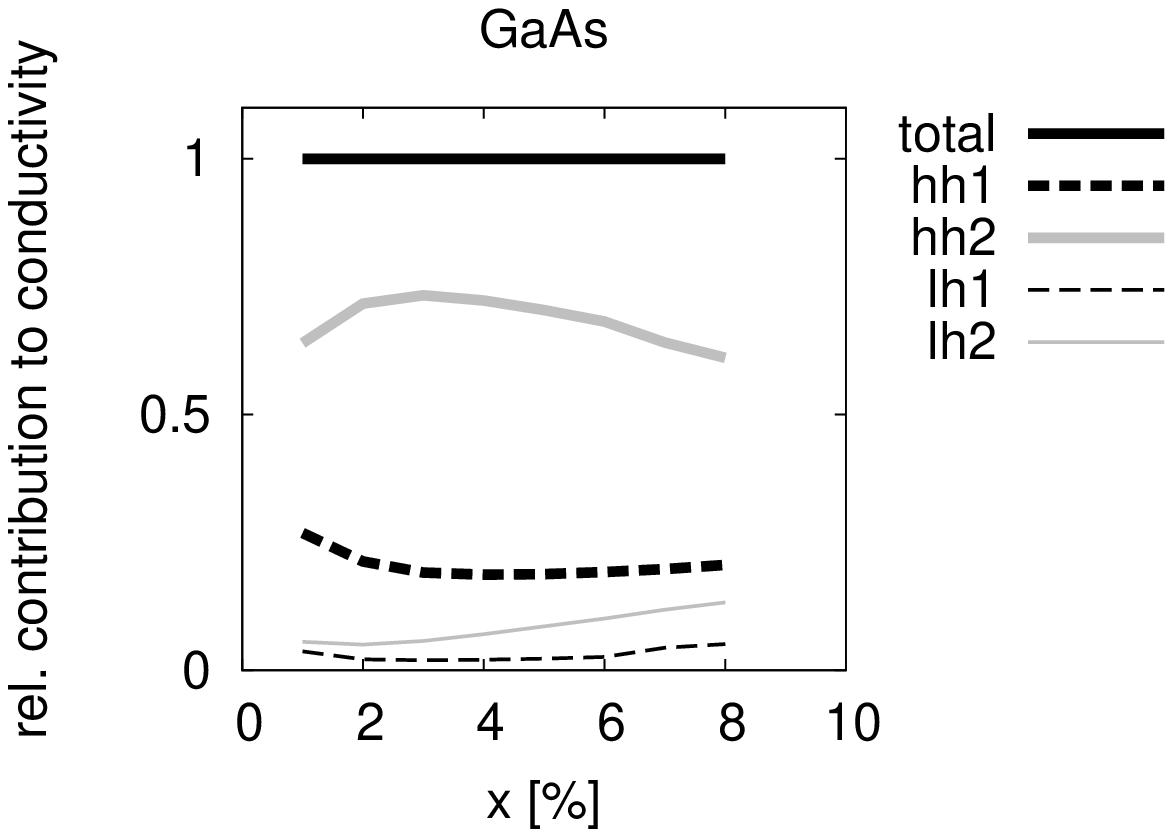} \\
  (a) & (b)
\end{tabular}
\end{center}
\caption{Contribution of particular bands to the calculated 
  total bulk conductivity in
  uncompensated GaAs. (a) Absolute values, (b) relative to the total
  conductivity.}
\label{fig-01}
\end{figure}

\begin{figure}
\begin{center}
\begin{tabular}{c}
\hskip-2cm
\includegraphics{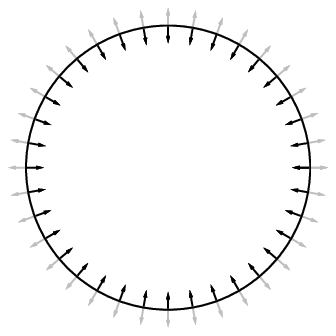}\hskip-1cm\raise5mm\hbox{%
\includegraphics{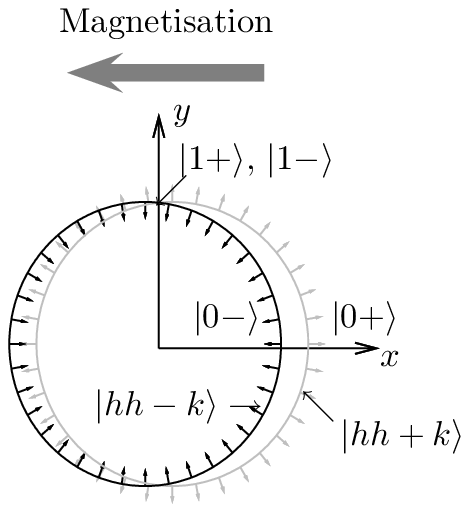}}
\end{tabular}
\end{center}
\caption{Heavy hole bands. Spin texture without and with exchange
  splitting. }
\label{fig-02}
\end{figure}

Numerically, the light hole bands are found to give only small contribution to
the total conductivity as shown in Fig.~(\ref{fig-01}), hence we consider only
the heavy hole bands. This is a substantial simplification since the
Hamiltonian (\ref{eq-02}) restricted to the heavy-hole subspace
$j_k=\pm\frac{3}{2}$ can be diagonalized analytically (see Appendix A). By
denoting the angle between $\vec{k}$ and $\hat{M}$ as $\phi_{\vec{k}}$ the
eigenenergies read
\begin{equation}\label{eq-07}
  E_{1,2} = \textstyle
    (\gamma_1-2\gamma_2) \hbar^2 k^2/2m 
    \pm h \cos \phi_{\vec{k}}\,.
\end{equation}
In the presence of the exchange field $h$ the two originally identical
heavy-hole Fermi spheres transform into (typically weakly) distorted '$+$' and
'$-$' spheres displaced by $\Delta k=\pm (2m_*/\hbar)^{1/2} h/4\sqrt{E_F}$
along $\hat{M}$.  The eigenstates $|hh+\vec{k}\rangle$, $|hh-\vec{k}\rangle$
are independent of $h$ and correspond to perfectly isotropic radial spin
textures: expectation value of $\vec{s}$ in the state $|hh+\vec{k}\rangle$
($|hh-\vec{k}\rangle$) is a vector of length $\frac{1}{2}$ parallel to
$\vec{k}$ and pointing outwards (inwards), Fig.~\ref{fig-02}. We remark that
Eq.~(\ref{eq-07}) and $|hh\pm \vec{k}\rangle$ remain unchanged even when the
Hamiltonian~(\ref{eq-07}) is extended to include a finite $\Delta_{SO}$ (see
Appendix A).

In the lowest order Born approximation, the
transport relaxation time is given by
\begin{equation}\label{eq-03}
  \frac{1}{\tau(\vec{k})} = \int \frac{d^3\vec{k}'}{(2\pi)^3} \frac{2\pi}{\hbar} 
  N_{\mathrm{Mn}}
  \delta(E(\vec{k})-E(\vec{k}')) |M_{\vec{k}\vec{k}'}|^2 (1-\hat{k}\cdot\hat{k}')\,.
\end{equation}
The last factor corresponds to vertex corrections in the Kubo formalism and
$\hat{k}\cdot\hat{k}'$ is the cosine of the angle between $\vec{k}$ and
$\vec{k}'$.  The scattering matrix element between initial
$|z_{\vec{k}'}\rangle$ and final $|z_{\vec{k}}\rangle$ states is
\begin{equation}\label{eq-08}
  M_{\vec{k}\vec{k}'}^C = \langle z_{\vec{k}}| V(\vec{k}-\vec{k}') |z_{\vec{k}'}\rangle = 
  V(|\vec{k}-\vec{k}'|)
  \langle z_{\vec{k}} | z_{\vec{k}'}\rangle
\end{equation}
for the isotropic (e.g. screened Coulomb) scattering off a charged ion,
and
\begin{equation}\label{eq-09}
  M_{\vec{k}\vec{k}'}^B = \langle z_{\vec{k}}| (h/N_{\mathrm{Mn}}) 
  \hat{M}\cdot \vec{s} |z_{\vec{k}'}\rangle =
  J_{pd}S_{\mathrm{Mn}} \langle z_{\vec{k}}| \hat{M}\cdot \vec{s} |z_{\vec{k}'}\rangle
\end{equation}
for the scattering off a magnetic moment. These
two add up coherently for substitutional Mn so that
$M_{\vec{k}\vec{k}'}=M_{\vec{k}\vec{k}'}^B+M_{\vec{k}\vec{k}'}^C$.

To simplify the further qualitative analytical discussion we replace the
long-range Coulomb potential with an effective $\delta$-function potential and
denote the ratio of this non-magnetic potential and of $J_{pd}S_{Mn}$ as
$\alpha$. Our relaxation rates will be evaluated from the following scattering
operator
\begin{equation}\label{eq01}
  M_{\vec{k}\vec{k}'}\propto 
     \langle z_{\vec{k}}| \alpha + s_x|z_{\vec{k}'}\rangle\,.
\end{equation}
We discarded an overall prefactor, since when only one source of scattering is
present, this prefactor will cancel in the expressions for relative changes of
conductivity needed to obtain the AMR. 

Equation (\ref{eq-03}) will now be used to calculate the scattering rates for
two special values of $\vec{k}$, details of that calculation are given in
Appendix B. The first one is $\vec{k}||\hat{M}=\hat{e}_x$.  The corresponding
initial states for the scattering, $|hh\pm \vec{k}\rangle$ will be abbreviated
as $|0\pm\rangle$, see Fig.~\ref{fig-02}. Integrands in Eq.~(\ref{eq-03})
can be simplified considerably noting that $s_x|0\pm\rangle =
\pm\frac{1}{2}|0\pm\rangle$. We obtain 
\begin{eqnarray}\label{eq-05}
  \frac{1}{\tau_\pm^x} &=& \int_{FS} d^2\vec{k}' \left\{
   |\langle hh+\vec{k}'|0\pm\rangle|^2(\alpha\pm\frac{1}{2})^2 + \right. \\
\nonumber
   &&
   \left.+
|\langle hh-\vec{k}'|0\pm\rangle|^2(\alpha\pm\frac{1}{2})^2
   \right\}
   (1-\hat{k}\cdot\hat{k}')\,.
\end{eqnarray}
The integration variable is taken as dimensionless and the result should be
multiplied by a factor $R=2mk_F/h^2\cdot 2\pi/\hbar\cdot N_{\mathrm{Mn}}\cdot
(J_{pd}S_{\mathrm{Mn}})^2$ to get the real inverse scattering times based on
Eqs.~(\ref{eq-08},\ref{eq-09}) with $V(\vec{k}-\vec{k}')\equiv \alpha
J_{pd}S_{\mathrm{Mn}}$.  By integrating Eq.~(\ref{eq-05}) we obtain,
\begin{eqnarray*}
  \textstyle
  1/\tau_+^{x}=2\pi(\alpha-\frac{1}{2})^2 &\qquad & 
  \textstyle
  1/\tau_-^{x}=2\pi(\alpha+\frac{1}{2})^2\,.
\end{eqnarray*}
In contrast, the other pair of initial states $|1\pm\rangle$
with $\vec{k}\perp \hat{M}$ (see Fig.~\ref{fig-02}), gives
\begin{eqnarray} \nonumber
  \frac{1}{\tau_\pm^y} &=& \int_{FS} d^2\vec{k}' \left\{
   |\langle hh+\vec{k}'|1\pm\rangle|^2 \alpha^2 +
   |\langle hh+\vec{k}'|s_x|1\pm\rangle|^2 +\right. \\
  && \label{eq-06}
   \hskip-1.2cm \left.
   |\langle hh-\vec{k}'|1\pm\rangle|^2 \alpha^2 +
   |\langle hh-\vec{k}'|s_x|1\pm\rangle|^2 +
   \right\}
   (1-\hat{k}\cdot\hat{k}')\,.
\end{eqnarray}
The integrals not involving $s_x$ give again $2\pi$ by the virtue of symmetry
(cf. Eq.~(\ref{eq-05}) and Fig.~\ref{fig-02}) while the remaining two terms
have to be evaluated independently to give
\begin{eqnarray}\label{Eq14}
& 1/\tau_+^y=1/\tau_-^y=2\pi(\alpha^2+\frac{1}{12})& 
\end{eqnarray}
Conductivities for $\hat{M}||I$ and $\hat{M}\perp I$ can be estimated based on
these relaxation rates as $\sigma_{||}=P(\tau_+^x+\tau_-^x)$ and  
$\sigma_{\perp}=P(\tau_+^y+\tau_-^y)$, i.e.
\begin{eqnarray}\label{Eq15}
  \sigma_{||}/P=
    \frac{1}{(\alpha-\frac{1}{2})^2}+\frac{1}{(\alpha+\frac{1}{2})^2}
  &\qquad&
  \sigma_{\perp}/P = \frac{2}{\alpha^2+\frac{1}{12}}\,,
\end{eqnarray}
where $P=e^2v_F^2/3\cdot g R/2\pi$, $v_F$ is the Fermi velocity and $g$ is the
density of states at the Fermi wavevector $k_F$. For heavy holes, 
$v_F=(\gamma_1-2\gamma_2)\hbar k_F/m$ and $g=2mk_F/h^2(\gamma_1-2\gamma_2)$.
Eq.~(\ref{Eq15}) leads to
\begin{equation}
  \frac{\sigma_{||}}{\sigma_\perp} =
  \frac{(\alpha^2+\frac{1}{12})(\alpha^2+\frac{1}{4})}{
  (\alpha^2-\frac{1}{4})^2}\,,
\end{equation}
presented in Ref.~\cite{Rushforth:2007}. The AMR, using the notation of the
experimental section %(see Fig.~\ref{fig-03}), 
then reads
\begin{equation}
  \mathrm{AMR} = 2\frac{\rho_{xx}(\hat{M}|| I)-\rho_{xx}(\hat{M}\perp I)}
{\rho_{xx}(\hat{M}|| I)+\rho_{xx}(\hat{M}\perp I)}=
  -2\frac{\sigma_{||}-\sigma_{\perp}}{\sigma_{||}+\sigma_\perp} =
  -\frac{20\alpha^2-1}{24\alpha^4 - 2\alpha^2 + 1}\,.
\end{equation}

\begin{figure}
\begin{tabular}{c}
  \includegraphics[scale=0.5]{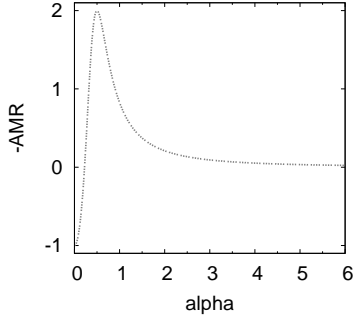} 
\end{tabular}
\caption{AMR as a function of the ratio between effective strengths of the
  non-magnetic and magnetic scattering.}
\label{fig-03}
\end{figure}

When the magnetic term in the impurity potential is much larger than the
non-magnetic term ($\alpha\ll 1$) one expects $\sigma_\parallel<\sigma_\perp$ 
(positive AMR, as is usually observed in
metallic ferromagnets).  However, 
the sign of the non-crystalline AMR reverses at a
relatively weak non-magnetic potential ($\alpha=1/\sqrt{20}$ in the model),
its magnitude is then maximized when the two terms are comparable
($\alpha=1/2$), and, for this mechanism, it vanishes when the magnetic term is
much weaker than the non-magnetic term ($\alpha\rightarrow\infty$).
Note that the large magnitude of the AMR for $\alpha=1/2$ due to
$1/\sigma_{\parallel}=0$ is an artefact of the special form of the simplified scattering operator (9) where the magnetic and non-magnetic part depend in exactly the same way on $k$ (they are constant, i.e. they both correspond to point-like scatterers).

%an artefact of the approximation calculating the
%relaxation rates only for $k$-states parallel to the current direction.

Physically, carriers moving along ${\hat{M}}$, {\em i.e.} with $\vec{s}$
parallel or antiparallel to ${\hat{M}}$, experience the strongest
scattering potential among all Fermi surface states when $\alpha=0$, giving
$\sigma_{\parallel}<\sigma_{\perp}$. When the
non-magnetic potential is present, however, it can more efficiently cancel the
magnetic term for carriers moving along $\hat{M}$, and for relatively
small $\alpha$ the sign of AMR flips. Since $\alpha<1/\sqrt{20}$ is
unrealistic for the magnetic acceptor Mn in GaAs
\cite{Jungwirth:2002_c,Jungwirth:2006_a} we obtain 
$\sigma_{\parallel}>\sigma_{\perp}$, consistent with experiment.

\section{Summary}

A phenomenological framework for the AMR in the zinc-blende crystalline
environment of GaMnAs was used to analyse experimental data from (i) bulk
material, (ii) thin layers, and (iii) samples with lithographically
manipulated strain. While leaving the crystalline components aside for a
further theoretical study, a qualitative analytical model was presented for
the non-crystalline AMR. The model is based on the anisotropic relaxation
times in the heavy hole bands due to a combined magnetic/non-magnetic
scattering on Mn impurities.  It offers an explanation of the sign of the
non-crystalline AMR in GaMnAs which is opposite to most of the conventional
metal ferromagnets.

We acknowledge support from EU Grant IST-015728, from UK Grant GR/S81407/01,
from CR Grants 202/05/0575, 202/04/1519, FON/06/E002, AV0Z1010052,
KAN400100652, and LC510, from ONR Grant N000140610122, and from SWAN. 
J.~Sinova is a Cottrell Scholar of Research Corporation.

% The Appendices part is started with the command \appendix;
% appendix sections are then done as normal sections

\appendix

\section{Some basic algebra with Kohn--Luttinger Hamiltonian}

Equation~(\ref{eq-07}) 
and explicit form of the heavy-hole eigenvectors $|hh\pm
\vec{k}\rangle$ will be derived in this Appendix. As we will refer to the
spherical approximation with $\hat{M}||\hat{e}_x$, the main body of
calculations will be carried out in the $k_x,k_y$ plane keeping in mind the
rotational symmetry of the problem around $\hat{e}_x$.  For the sake of
completeness, results for general $k_z\not=0$ will be given at the end of each
part without derivation.

The starting point is $H_{KL}$ as in Eq.~(A8) of 
Abolfath~{\em et~al.}\cite{Abolfath:2001_a} 
\begin{equation}\label{eq-A00}
  H_{KL}=\left(\begin{array}{cccccc}
      H_{hh} & -c & -b & 0 & b/\sqrt{2} & c\sqrt{2} \\
      -c^* & H_{lh} & 0 & b& -b^*\sqrt{3}/\sqrt{2} & -d \\
      -b^* & 0 & H_{lh} & -c & d & -b\sqrt{3}/\sqrt{2}\\
      0 & b^* & -c^* & H_{hh}& -c^*\sqrt{2} & b^*/\sqrt{2} \\
      b^*/\sqrt{2} & -b\sqrt{3}/\sqrt{2}& d^* & -c\sqrt{2} & H_{so} & 0 \\
      c^*\sqrt{2} & -d^* & -b^*\sqrt{3}/\sqrt{2}& b/\sqrt{2} & 0 & H_{so}
    \end{array}\right)\,.
\end{equation}
Going into the $k_x,k_y$ plane by putting $\vec{k}=k(\cos\phi,\sin\phi,0)$,
the off-diagonal elements become
$$
  b=0\,,\qquad
  c=\frac{\sqrt{3}\hbar^2k^2}{2m}[\gamma_2\cos2\phi - i\gamma_3\sin2\phi]\,,
  \qquad
  d=\frac{\sqrt{2}\hbar^2k^2}{2m}\gamma_2\,,
$$
and the diagonal elements are independent of $\phi$
$$
  H_{hh}=\frac{\hbar^2k^2}{2m}(\gamma_1+\gamma_2)\,,
  \qquad
  H_{lh}=\frac{\hbar^2k^2}{2m}(\gamma_1-\gamma_2)\,,
  \qquad
  H_{so}=\frac{\hbar^2k^2}{2m}\gamma_1 + \Delta_{SO}\,.
$$
It is convenient to factor out $\hbar^2 k^2/(2m)$ and to introduce
$\delta=\Delta_{SO}/(\gamma_2 \hbar^2/2m)\cdot k^{-2}$.

The spherical approximation relies in setting $\gamma_2=\gamma_3$ and allows
for the simplification
$$
  c/(\hbar^2 k^2/(2m))=\sqrt{3}\gamma_2\exp(-2i\phi)\,.
$$
In this approximation and even with general $k_z\not=0$,
i.e. 
\begin{equation}\label{generalK}
\vec{k}=k(\sin\theta\cos\phi, \sin\theta\sin\phi, \cos\theta)
\end{equation}
the spectrum of the Hamiltonian~(\ref{eq-A00}) does not depend on
$\theta,\phi$. Below, this statement is demonstrated explicitly for
$\theta=\pi/2$.

\subsection{Spectrum}

When $k_z=0$, rows/columns 3,4,5 of $H_{KL}$ are totally decoupled from
rows/columns 1,2,6. The $6\times 6$ problem is reduced to two independent and
equivalent $3\times 3$ problems. The latter (lines 1,2,6) is
\begin{equation}\label{eq-A01}
  A_+=H/(\hbar^2k^2/2m)=\left(\begin{array}{ccc}
      \gamma_1+\gamma_2 & -c & c\sqrt{2} \\
      -c^* & \gamma_1-\gamma_2 & -\sqrt{2}\gamma_2 \\
      \sqrt{2}c^* & -\sqrt{2}\gamma_2 & \gamma_1+\gamma_2\delta
      \end{array}\right)\,.
\end{equation}
Matrix $A_+$ depends on $\phi$ only via $c$. In the spherical approximation, 
$cc^*=|c|^2=3\gamma_2^2$ is $\phi$-independent.
The eigenvalues of $A_+$ are calculated from $\det(A_+-\lambda)=0$. This
determinant will depend on $c$ only via the combination $cc^*$ and hence it is
$\phi$-independent for $\gamma_2=\gamma_3$. 

\begin{figure}
\begin{tabular}{c}
  \includegraphics[scale=0.5]{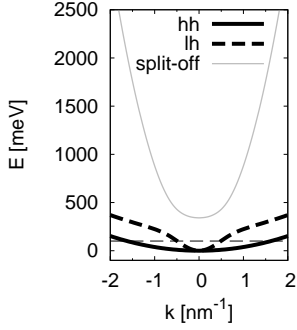} 
\end{tabular}
\caption{Dispersions of the heavy holes, light holes and the split-off band in
the spherical approximation. Horizontal dashed line indicates the Fermi energy
of $E_F=$100meV.}
\label{fig-04}
\end{figure}

Explicitly, Fig.~\ref{fig-04}, the eigenvalues of $A_+$, multiplied back by
$\hbar^2k^2/2m$, are
\begin{eqnarray}\nonumber
  E_1&=& \frac{\hbar^2k^2}{2m}\left[\gamma_1-2\gamma_2\right] \\
\label{eq-A05}
  E_2&=& \frac{\hbar^2k^2}{2m}\left[\gamma_1+\gamma_2\cdot 
         \frac{1}{2}(\delta+2-\sqrt{\delta^2-4\delta+36})\right]\\
\nonumber
  E_3&=& \frac{\hbar^2k^2}{2m}\left[\gamma_1+\gamma_2\cdot 
         \frac{1}{2}(\delta+2+\sqrt{\delta^2-4\delta+36})\right]\,.
\end{eqnarray}
They correspond to heavy holes, light holes and the split-off band, the last
two depend on $k$ via $\delta$, but none of them depends on $\phi$ and hence
give circular FS sections. The light-hole energy is
$[\gamma_1+2\gamma_2]\hbar^2k^2/(2m)$ in the limit $\Delta_{SO}\to \infty$.

Analysis of the complete Hamiltonian $H_{KL}$ is completed by constructing
matrix $A_-$ from lines 3,4,5. It is not identical to $A_+$, but it has an
identical spectrum to $A_+$. Thus, the Fermi surfaces of $H_{KL}$ in the
$k_z=0$ sections comprise of, largest to smallest, two coincident circles for
heavy holes with $E_F= (\gamma_1-2\gamma_2)\hbar^2k_F^2/2m$, which is
independent of $\Delta_{SO}$, two coincident circles for light holes and two
coincident circles for split-off bands.

\subsection{Eigenvectors}

Even though the spectrum of the Hamiltonians (\ref{eq-A00}) or (\ref{eq-A01})
does not depend on $\phi$, $\theta$, the eigenvectors do. 

The lowest-energy ($E_1=\gamma_1-2\gamma_2$) eigenvector of $A_+$ is
independent of $\delta$. Together with its counterpart from $A_-$ they
represent the degenerate heavy-hole states of $H_{KL}$ regardless of
$\Delta_{SO}$:
$$
  \vec{v}_{1+}=\frac{1}{2}(e^{-2i\phi},\sqrt{3},0,0,0,0)^T\,,\qquad
  \vec{v}_{1-}=\frac{1}{2}(0,0,\sqrt{3},e^{2i\phi},0,0)^T\,. 
$$
The component notation refers to the basis~(A5) of
Ref.~\cite{Abolfath:2001_a}, the superscript denotes the transposition. For
a general $\vec{k}$ given by Eq.~(\ref{generalK}) the heavy-hole states are
$$
  \vec{v}_{1+}=\frac{1}{2}(e^{-2i\phi}\sin\theta ,\sqrt{3}\sin\theta ,
  0,- e^{i\phi}\cos\theta,0,0)^T\,,\qquad
$$
$$
  \vec{v}_{1-}=\frac{1}{2}(e^{-i\phi}\cos\theta ,0,
  \sqrt{3}\sin\theta,e^{2i\phi}\sin\theta,0,0)^T\,. 
$$

\subsection{Perturbation theory with kinetic exchange}

The $p$-$d$ kinetic exchange has the form of an effective Zeeman field in the
magnetisation direction which we choose to be $x$, Fig.~\ref{fig-02}. The
corresponding operator (with respect to the first four basis vectors of (A5)
in~Ref.~\cite{Abolfath:2001_a}), that is the last term in Eq.~(\ref{eq-02}), is
\begin{equation}\label{eq-A02}
  H_{pd} = hs_x = h\frac{1}{6}\left(\begin{array}{cccc}
  0 & 0 & \sqrt{3} & 0 \\
  0 & 0 & 2 & \sqrt{3} \\
  \sqrt{3} & 2 & 0 & 0 \\
  0 & \sqrt{3} & 0 & 0 
  \end{array}\right)\,.
\end{equation}
The matrix of $H_{pd}/h$ in the basis of $\vec{v}_{1+}$, $\vec{v}_{1-}$ and
its diagonalization results are the following
$$
  H_{pd}/h = \frac{1}{4}\left(\begin{array}{cc} 
      0 & 1+e^{2i\phi} \\ 
      1+e^{-2i\phi} & 0
      \end{array}\right)\,,\qquad    
    \begin{array}{l}
    E_+= \frac{1}{2}\cos\phi\,,\quad \vec{v}_+ = 2^{-1/2} (1, e^{-i\phi})^T \\
    E_-=-\frac{1}{2}\cos\phi\,,\quad \vec{v}_- = 2^{-1/2} (1, -e^{-i\phi})^T\,.
    \end{array}                  
$$
In the limit of small $h$ ('degenerate-level perturbation calculus') the
degenerate heavy-hole bands of $H_{KL}$ become split in energy by $h\cos
\phi$. Recast into the $\vec{k}$-plane, the two coincident Fermi circles
become displaced. Their wavefunctions are
\begin{eqnarray}\label{eq-A03}
  |hh+,\vec{k}\rangle  &=&\frac{1}{2\sqrt{2}}
    (e^{-2i\phi},\sqrt{3}, \sqrt{3}e^{-i\phi}, e^{i\phi},0,0)^T\,,\\
  |hh-,\vec{k}\rangle  &=&\frac{1}{2\sqrt{2}}
    (e^{-2i\phi},\sqrt{3},-\sqrt{3}e^{-i\phi},-e^{i\phi},0,0)^T\,. \nonumber
\end{eqnarray}
For general $k_z\not=0$ (\ref{generalK}) the energies are
$$
  E_\pm = \pm \frac{1}{2} \cos\phi\sin\theta\,,
$$
as stated in Eq.~(\ref{eq-07}),
and the eigenvectors
\begin{eqnarray}\label{eq-A04}
  |hh+,\vec{k}\rangle  &=&
 (e^{-2i\phi}C^3,\sqrt{3} CS^2, \sqrt{3}e^{-i\phi} C^2S, e^{i\phi}S^3,0,0)^T\,,
\\ \nonumber
  |hh-,\vec{k}\rangle  &=&
 (e^{-2i\phi}S^3,\sqrt{3} SC^2, -\sqrt{3}e^{-i\phi} S^2C,-e^{i\phi}C^3,0,0)^T\,
\end{eqnarray}
with $C=\cos\theta/2$ and $S=\sin\theta/2$.

\section{Integrals of overlaps, Eqs. (\ref{eq-05},\ref{eq-06})}

The overlaps of eigenvectors in Eq.~(\ref{eq-A03}) are
\begin{eqnarray*}
  \langle hh+\vec{k}|hh+\vec{k}'\rangle &=& \frac{1}{8}[(e^{2i(\phi-\phi')}+3)+
                    e^{i(\phi-\phi')}(e^{-2i(\phi-\phi')}+3)]\,,\\
  \langle hh-\vec{k}|hh+\vec{k}'\rangle &=& \frac{1}{8}[(e^{2i(\phi-\phi')}+3)-
                    e^{i(\phi-\phi')}(e^{-2i(\phi-\phi')}+3)]\,.
\end{eqnarray*}
Note that these overlaps for $J=\frac{3}{2}$ spinors are not the same as for
$J=\frac{1}{2}$ spinors $|a\pm\vec{k}\rangle = (1,\pm e^{i\phi})^T$ even
though the spin textures of $|hh\pm\vec{k}\rangle$ and $|a\pm\vec{k}\rangle$
are the same, Fig.~\ref{fig-02}.

The two summands of integral in Eq.~(\ref{eq-05}),
\begin{eqnarray}\label{eq-C01}
  \int_{FS} d^2\vec{k}' |\langle hh + \vec{k}'|0\pm\rangle|^2
   (1-\hat{k}\cdot\hat{k}')
\end{eqnarray}
will be integrated with spherical coordinates $\phi',\Phi'$ with 'north pole'
$\phi'=0$ in $\vec{k}'||\hat{e}_x$. The polar angle ($\Phi'$) integration 
gives a factor of $2\pi$ and we obtain
$$
  \frac{2\pi}{64}\int_0^\pi d\phi' \sin\phi' \cdot
  2[10\pm 15\cos\phi' + 
  6\cos 2\phi' \pm \cos 3\phi'] (1-\cos\phi')
$$
For the $+$ sign we get $4\pi(6+\frac{2}{5})/64$ and for $-$ it is
$4\pi(26-\frac{2}{5})/64$ giving the total of $2\pi$. The integrals with
$\langle hh-\vec{k}'|$ in Eq.~(\ref{eq-C01}) are analogous.

Concerning the integral in Eq.~(\ref{eq-06}), we first have to show that it is
indeed equal to the scattering rate
\begin{eqnarray}\nonumber
  \frac{1}{\tau_\pm^y} &=&  \int_{FS}  d^2\vec{k}' \left\{
   |\langle hh+\vec{k}'|\alpha+s_x |1\pm\rangle|^2 + \right. \\
   &&\left.|\langle hh+\vec{k}'|\alpha+s_x \nonumber
  |1\pm\rangle|^2\right\}(1-\hat{k}\cdot \hat{k}')\,.
\end{eqnarray}
In order to show this, it suffices to demonstrate that 
$$
  \int_0^{2\pi} d\Phi' \langle hh+\vec{k}'|\alpha|1+\rangle 
                       \langle 1+|s_x|hh+\vec{k}'\rangle =0\,
$$
which holds by the virtue of
$\langle hh+\vec{k}'|s_x|1+\rangle\propto e^{i\Phi'}$
and $\langle hh+\vec{k}'|1+\rangle\propto e^{2i\Phi'}$ \cite{phaseNote}.

Proceeding with  Eq.~(\ref{eq-06}), we will use
\begin{eqnarray*}
  \langle hh+\vec{k}|s_x|hh+\vec{k}'\rangle &=&
     \frac{1}{8}(e^{i\phi}+e^{-i\phi'})[1+\cos(\phi-\phi')] \\
  \langle hh-\vec{k}|s_x|hh-\vec{k}'\rangle &=&
     - \langle hh+\vec{k}|s_x|hh+\vec{k}'\rangle\\
  \langle hh-\vec{k}|s_x|hh+\vec{k}'\rangle &=&
     -\frac{1}{8}(e^{i\phi}-e^{-i\phi'})[1-\cos(\phi-\phi')] \\
  \langle hh+\vec{k}|s_x|hh-\vec{k}'\rangle &=&
     -\langle hh-\vec{k}|s_x|hh+\vec{k}'\rangle
\end{eqnarray*}
Here, the proper choice of $\vec{k}$ ($||\hat{e}_y$) 
amounts to putting $\phi=\frac{\pi}{2}$ and the
integral for scattering from $|1+\rangle$ to the $+$ band is
\begin{eqnarray*}
  \frac{2\pi}{64}\int_{-\pi/2}^{\pi/2} d\phi' \cos\phi' 
  [1+\sin\phi']^2 2(1-\sin\phi')\cdot (1-\sin\phi') = \frac{2\pi}{32} \cdot
  \frac{13}{120}\,,
\end{eqnarray*}
while the other integral ($\to -$) 
\begin{eqnarray*}
  \frac{2\pi}{64}\int_{-\pi/2}^{\pi/2} d\phi' \cos\phi' 
  [1-\sin\phi']^2 2(1+\sin\phi')\cdot (1-\sin\phi') = \frac{2\pi}{32} \cdot
  \left(\frac{8}{3}-\frac{13}{120}\right)\,.
\end{eqnarray*}
Their sum is $2\pi\frac{1}{12}$ which completes the proof of Eq.~(\ref{Eq14}).


\begin{thebibliography}{04}
\bibitem{Corbino}Strictly, in a Corbino geometry we measure
  $G^{C}=I/V=(\sigma_{xx}$+$\sigma_{yy})/2$ and plot $R^{C}=1/G^{C}$. If we
  note that $\rho_{xy}\ll\rho_{xx}$ to a good approximation $\Delta R^{C}/R_{av}^{C}=(\Delta
  R_{[110]}^{H}+\Delta R_{[1\bar{1}0]}^{H})/2R_{av}$=
  $C_{U}\cos2\psi+C_{C}\cos4\psi$ where the superscipts denote measurements on Hall bars or Corbino devices.
\bibitem{phaseNote} To see this, it is easier to take $|1+\rangle$ as
  $|hh+\vec{k}\rangle$ with $\vec{k}||\hat{e}_z$, 
  that is $\langle 1+|\propto(1,0,0,0)$. 
  Then $\langle 1+|s_x\propto (0,0,1,0)$ and
  using Eq.~(\ref{eq-A04}) $\langle 1+|s_x|hh+\vec{k}\rangle\propto
  e^{i\phi}$. On the other hand, again from Eq.~(\ref{eq-A04}), 
  $\langle 1+|hh+\vec{k}\rangle\propto e^{-2i\phi}$.
\end{thebibliography}
\end{document}